\documentclass{IEEEtran4PSCC}
\ifCLASSINFOpdf
\else
\fi
\usepackage[cmex10]{amsmath}
\usepackage{eurosym}

\usepackage{amssymb}
\usepackage{graphicx}
\usepackage{color}

\DeclareMathOperator*{\argmin}{arg\,min}

\newcommand{\bs}[1]{\boldsymbol{#1}}

\newcommand{\act}{\bs{u}}
\newcommand{\scenario}[1]{Z_{\act_{#1},\theta}}

\newcommand{\actionm}{u_m}

\newcommand{\actions}{u_p}

\newcommand{\actm}{u_m}

\newcommand{\s}[1]{S_{#1}}

\newcommand{\actionSetm}{\mathcal{U}_m}

\newcommand{\prob}[1]{\mathbb{P}\left\{#1\right\}}

\newcommand{\eq}[1]{Eq.~(\ref{#1})}
\newcommand{\eqs}[2]{Eqs.~(\ref{#1})-(\ref{#2})}

\makeatletter
\newcounter{parentalgorithm}

\usepackage{textpos}
\usepackage{algorithm,algorithmic}
\usepackage{makecell}

\newcommand{\Rmnum}[1]{\expandafter\@slowromancap\romannumeral #1@}
\makeatother

\begin{document}
\title{Distributed Scenario-Based Optimization for Asset Management in a Hierarchical Decision Making Environment}

%\title{Risk Minimization for Asset Management Scheduling  through Sampling}

%% To specify the authors when (number of affiliations <= 2)
\author{
	\IEEEauthorblockN{Gal Dalal, Elad Gilboa, Shie Mannor}
	\IEEEauthorblockA{Department of Electrical Engineering 
		Technion\\
		Haifa, Israel\\
		gald@tx.technion.ac.il, egilboa@tx.technion.ac.il,	shie@ee.technion.ac.il
}
}
%% To specify the authors when (number of affiliations > 2)
% \author{\IEEEauthorblockN{Author n.1\IEEEauthorrefmark{1},
% Author n.2\IEEEauthorrefmark{2},
% Author n.3\IEEEauthorrefmark{3}, 
% Author n.4\IEEEauthorrefmark{3} and
% Author n.5\IEEEauthorrefmark{4}}
% \IEEEauthorblockA{\IEEEauthorrefmark{1} Department Name of Organization A\\
% Name of the organization A,
% Address A\\ Emails if wanted}
% \IEEEauthorblockA{\IEEEauthorrefmark{2} Department Name of Organization B\\
% Name of the organization B,
% Address B\\ Emails if wanted}
% \IEEEauthorblockA{\IEEEauthorrefmark{3} Department Name of Organization C\\
% Name of the organization C,
% Address C\\ Emails if wanted}
% \IEEEauthorblockA{\IEEEauthorrefmark{4}Department Name of Organization D\\
% Name of the organization D,
% Address D\\ Emails if wanted}
% }

\maketitle


\begin{abstract}
	 Asset management attempts to keep the power system in working conditions. It requires much coordination between multiple entities and long term planning often months in advance. In this work we introduce a mid-term asset management formulation as a stochastic optimization problem, that includes three hierarchical layers of decision making, namely the mid-term, short-term and real-time. We devise a tractable scenario approximation technique for efficiently assessing the complex implications a maintenance schedule inflicts on a power system. This is done using efficient Monte-Carlo simulations that trade-off between accuracy and tractability. We then present our implementation of a distributed scenario-based optimization algorithm for solving our formulation, and use an updated PJM 5-bus system to show a solution that is cheaper than other maintenance heuristics that are likely to be considered by TSOs. 
\end{abstract}


% Use this to place sponsorships
\thanksto{The research leading to these results has received funding from the European Union Seventh Framework Programme (FP7/2007-2013) under grant agreement No 608540, project acronym GARPUR. The scientific responsibility rests with the authors. We wish to thank the following people for their helpful comments on this work: Prof. Louis Wehenkel, ULg; Efthymios Karangelos, ULg; Remy Clement, RTE.}

\section{Introduction}	\label{chp:intro}

Asset management is done by transmission system operators  (TSOs) in order to keep the power grid in operation by routinely maintaining the assets under their management. Maintaining an asset improves its working condition and resiliency, which reduces the risk of failures. However, planning and scheduling of maintenance jobs (which asset to maintain and when) is a complex task since it must take into account constrained resources (working crews, hours, and budget), increased vulnerability of the grid to contingencies during maintenance, and the impact of the necessary scheduled outages on short term operations and system security. Maintenance schedules, which are often planned several months into the future, must also deal with the large uncertainties arising from operating in a changing environment where high impact rare events can result in unacceptable consequences. 
In this work, we present a framework for solving the mid-term (months to a few years) asset-management problem under uncertainty using stochastic optimization. We present a general framework for asset management that trades off the direct cost of performing maintenance with future possible costs, such as of energy not supplied to customers due to asset failure.

When planning for future outages to enable maintenance, a certain reliability criterion is attempted to be satisfied at all future times. In present day, the common practice among TSOs is to consider the deterministic N-1 reliability criterion. Probabilistic criteria are also being investigated \cite{karangelos2015probabilistic,GARPUR_2.1}. In order to make the system N-1 compliant months in advance, the asset management operator must assess whether each of the possible future scenarios (taking into account the short-term day-ahead planing decisions and real-time operation decisions) are N-1 secure. Since taking into account all possible realizations of future events is impractical, they must be approximated using sampled paths of future scenarios. 
%For a concrete example, consider a TSO attempting to assess whether a certain maintenance schedule will result in a N-1 secured grid for the next 6 months. To do so, the TSO might test a few representative scenarios, which are chosen deterministically by experts or drawn randomly from a distribution.  However, since only a very small subset of possible scenarios may be considered, the effect of the actual future scenario in reality might have been left out and the grid might not be N-1 secured at some point in the future. 
For this reason, devising a good sampling scheme which gives a rich, informational representation of possible future occurrences is important, as well as reiterating the decision process as new information becomes available.

\subsection{Literature Review} \label{sec:literature_review}
Much work has been done in the asset management literature \cite{GARPUR_5.1,abiri2013two,jiang2004risk,entso-e,fu2007security}. Current state-of-the-art in transmission asset management offers  three main approaches: time-based preventive maintenance, condition-based preventive maintenance, and reliability-centered preventive maintenance. The trade-off, that usually comes into play in the objective function, is between increasing the transmission equipment reliability via maintenance, and minimizing the effect of transmission equipment maintenance outage on socio-economic welfare while satisfying operating constraints.

In \cite{abiri2013two}, these two aspects of the trade-off are being considered simultaneously. The authors use a linearized Weibull probability to calculate the probability of asset failure in future scenarios. Their method is based on a two-stage optimization formulation. The first stage involves a midterm asset maintenance scheduler that explicitly considers the analytic term of the probability asset failure scenarios, and the solver iterates on all of these possible scenarios. For this stage, a coarse division to time blocks is done, where each time block is assigned with its designated asset maintenance actions. The second stage introduces a short-term maintenance scheduler with the N-1 reliability criterion that schedules the output of the mid-term maintenance scheduler in the short run for each of the time blocks, with fine constraints such as security limits. The mid-term and short-term stages are completely decoupled schemes to make the problem computationally tractable. 

In a different work, Yong Jiang et. al \cite{jiang2004risk} focus on cumulative risk reduction as the objective of asset maintenance optimization. They define two important terms: severity and risk. Severity is defined to be a quantity describing the bad effect of four possible outcomes of contingencies assessed using power-flow simulations: overloads, cascading overloads, low voltage, and voltage collapse. The risk is then defined to be the product of the probability of a contingency happening and its severity. Each maintenance action has its added contribution to risk reduction, which is initially negative during the actual maintenance (due to the forced outage), and positive afterwards (due to its reduction in probability of future contingencies). The paper also addresses the asset life model, with functional description of Weibull and Markov models, and their appropriate parameter estimation description. Despite its broad system perspective and important contribution, this work necessitates strong assumptions such as an additive structure of the risk function, and the knowledge of generation and load profiles a year-ahead for each hour. In addition, the hourly year-long trajectories that are used for optimization are problematic because they introduce high variance to the optimization algorithm.

%Other works focus on maintenance outage scheduling of pre-determined maintenance actions, without considering the possible outcomes of asset failures and their condition. 

%
%Mathematically, the proposed problem is a large-scale, mixed-integer, non-convex optimization problem. In general, the entire outage scheduling problem is very large and cannot be solved in a reasonable computation time. The primary contributors to the complex structure of mid-term planning of maintenance outage scheduling are generation unit commitment (UC), economic dispatch (ED), resource allocation and utilization, and transmission security.
%
%To overcome these technical barriers, a coordination strategy between the different optimization tasks is proposed in \cite{fu2007security}. Planning is being performed in order to provide coordination between optimization strategies, where an optimal objective while is sought alternating between them. The authors use a deterministic model, and account for outages due to maintenance on generators and lines. In their model they do not account for asset condition degradation, contingencies and different future scenarios.

%comes before the word strategies:
%: coordination between generation and transmission maintenance outages; coordination between mid-term maintenance outage and hourly security-constrained generation scheduling; coordination between mid-term allocation and short-term utilization of resources; coordination between short-term transmission security and optimal maintenance outage scheduling
\subsection{Novelty of the Framework}

The complex dependence between the multiple time-horizons and the high uncertainty of the longer term planning, makes the corresponding decision making problems challenging. Here, we build on the idea of a coordination problem between a hierarchy of three decision layers, corresponding respectively to the mid-term, short-term and real-time contexts of reliability management. To deal with complexity of the model we suggest a scenario-based optimization approach based on efficient simulations that trade-off between accuracy and tractability.

In essence, the novelty of our work is three-fold. Firstly, our model is designed to include all different time horizons in one detailed and complete set of mathematical relations and notations. Secondly, we enable the use of complicated models by utilizing a simulation-based stochastic optimization approach that does not rely on analytical solutions. Lastly, we suggest a tractable methodology for solving the optimization problem, based on distributed computing and an efficient scenario sampling method.

\section{Mathematical Problem Formulation} \label{sec:problem_formulation}
We formalize the mid-term optimization problem in \eq{eq:fullMidTermFormulation}, in which the TSO goal is to minimize future costs by adopting an optimal planned maintenance schedule $\actionm$, which also corresponds to maximizing the social welfare. The evaluation spans over a time horizon of $T$ hours. The decision variable in \eq{eq:fullMidTermFormulation} is the maintenance schedule $\actionm$ that is composed of a sequence of preventive maintenance actions with certain moments of activation chosen in the corresponding horizon. Denote the maintenance plan $\actionm \in \actionSetm = \{0,1\}^{n^l \times T_M}$ to be a binary matrix, where $T_M$($=\frac{T}{24\cdot 30}$) are monthly time indices and $n^l$ is the number of transmission lines in the system each entry, $\actionm(i_m,i_l)$, mentions whether transmission line $i_l$ is maintained during month $i_m$ or not. 
To lower the complexity of the problem we only consider asset management of transmission lines. Similar to \cite{abiri2013two}, we assume maintenance resets the effective age. Maintenance is done once a month and has known duration.

The objective in \eq{eq:midTermObj} is composed of the direct deterministic cost of maintenance actions $C_{\actionm}(i_l)$ on asset $i_l$ and an expected value $\mathbb{E}_{Z} C(\scenario{m})$ of stochastic indirect costs (e.g., load shedding, redispatch). The indirect costs are associated with the uncertain future conditions of the grid, denoted by $\scenario{}$ and under the shorter time decision policy (i.e., short-term operation planning and real-time) $\theta$ (e.g., $N-1$ security criterion). 
When making decisions in the mid-term time-horizon, one must take into account the shorter time decisions  that take place during these time intervals. The shorter time-horizons decisions include the short-term (day-ahead) operational planning decision $\actions \in \mathcal{U}_p(u_m)$ and real-time control $u_{RT} \in \mathcal{U}_{RT}(u_p)$ decision. Each of the sets of possible shorter time decisions $\mathcal{U}_p(u_m),~\mathcal{U}_{RT}(u_p)$ is defined by decisions that were taken one level higher in the hierarchy.

 The resulting formulation is therefore the following stochastic optimization problem:
%	We do this by initially attempting avoid load shedding completely by removing it from the decision, and only if no feasible solutions are available, we allow load-shedding
\begin{subequations}\label{eq:fullMidTermFormulation}
	\begin{align}
	& \min_{\actionm \in \actionSetm} \sum_{i_m=1}^{T_M} \sum_{i_l=1}^{n^l} C_{\actionm}(i_l)\actionm(i_m,i_l) \label{eq:midTermObj}  \\
	&\quad \quad \quad \quad +  \mathbb{E}_{Z} \left\{C(\scenario{m},\actionm, u_p^*, u_{RT}^*)\right\}  \nonumber\\
	& \text{subject to} \nonumber\\
	&h(\actionm) \le 0  \label{eq:midTermFeasibility}\\
	&u_p^* = \argmin_{u_p \in \mathcal{U}_p(u_m)}   C_{p}(y_m,u_m,u_p, u_{RT}^*) \label{eq:optimalS}\\
	& \quad\quad\quad ~ s.t \quad u_{RT}^* = \argmin_{u_{RT} \in \mathcal{U}_{RT}(u_p)}  C_{RT}(y_{RT},u_p, u_{RT}) 	\label{eq:optimalRT}
	\end{align} 
\end{subequations}
The reliability and cost of a power-system are defined in real-time. A real-time decision depends on the decisions taken in the short-term planning, which in turns depends on decision taken in the mid-term. Therefore 
\begin{align} \label{eq:scenario_cost}
&\mathbb{E}_{Z} \left\{C(\scenario{m},\actionm, u_p^*, u_{RT}^*)\right\}\\
&=\sum_{t=1}^{T} \mathbb{E}_{s_t \in Z} \left\{C_{RT}(\s{t},\actionm, u_p^*, u_{RT}^*)\right\} \nonumber
\end{align} is the expected cost of real-time operational decisions summed over the evaluation horizon, where the expectation is over the distribution of a stochastic scenario $\scenario{m}$ that is composed of a series of states $\s{t}$, as explained in \S ~\ref{sec:state-space}. 
Maintenance feasibility constraints $h(\actm)$ in \eq{eq:midTermFeasibility} define which maintenance actions are feasible, e.g., cannot maintain more than two assets per month.
The constraints in \eqs{eq:optimalS}{eq:optimalRT} describe the connection between the different time-horizon decisions, with short-term operational planning cost $C_p$ and real-time control cost $C_{RT}$,  where an optimal solution in one time-horizon must take into account how it will effect the future shorter time-horizon decisions. The informational states $y_m$ and $y_{RT}$ appearing as arguments in $C_p,~C_{RT}$ are revealed to the decision makers in these time-horizons, on which we expand in \S ~\ref{sec:informational_state}.

In this problem we prioritize high reliability by avoiding load shedding even when paying the cost of lost load (amount of load shedding times the value of lost load) is financially preferable. In \S ~\ref{sec:action_space} we explain how this is being done with our escalation process. In the following section we describe in detail each of the terms in this problem. 
\section{Probabilistic Model Components and Definitions}	
\label{sec:components}
In this section we present the definitions and  probabilistic mathematical model  used in this chapter. We define a state-space representation in terms of the states of the world in which the power system, the decision maker, and their exogenous environment can be in and their evolution in time (see Fig.~\ref{fig:scenarioDescription}). This is a generic model that can be adapted for the study of any transmission system by adjusting the definitions of the states, actions and transitions, as will be defined later. 
\subsection{State-Space}	
\label{sec:state-space}
We use a \textit{state} notation $\s{t}\in \mathcal{S}$ to represent all the information about the grid and its external environment at some time point $t$, needed to make informed decisions  \cite{puterman1990markov}. It includes the relevant information of all three time horizons, namely mid-term, short-term and real-time.
Denote $n^l_t,~n^b,~n^g_d,~n^g_w$ to be the number of transmission lines, number of buses, number of dispatchable generators, and number of wind generators in the network respectively. The topology $top_t$ changes with time and certain lines become unavailable when under maintenance, hence the time index in the variable $n^l_t$ which will be used from now on.  
The state $\s{t}$ is defined as the following tuple:
\[\s{t} = (\tau_t,\hat{W}_{d.a},\hat{D}_{d.a},W_t,D_t,top_t)\]
where
\begin{itemize}
	\item $\tau_t \in \mathbb{N}^{n^l}$ is the effective age of each of the assets at time $t$, which dictates their failure probability.
	\item $\hat{W}_{d.a} \in \mathbb{R}_+^{n^g_w \times T_{d.a}}$ is the day-ahead wind generation forecast. The rest of the $n^g_d$ dispatchable generators are fully controlled by the short-term and real-time decision makers, and therefore are deterministic (determined by the commitment plan $\actions$ and dispatch $\actions$) and are not a part of the state.
	%		, which is governed by the day-ahead unit-commitment plan of the $n^g_d$ dispatchable generators, and the wind power prediction of the $n^g_w$ generators in the system
	$T_{d.a}$ is the day-ahead planning horizon, taken to be 24 in our simulations. \\To avoid confusion, all variables with subscript $d.a$ do not include a time index subscript, rather they stay fixed for time periods of length $T_{d.a}$, and are updated each $T_{d.a}$ time-steps.
	\item $\hat{D}_{d.a} \in \mathbb{R}_+^{n^b \times T_{d.a}}$ is the day-ahead load forecast.
	%		\item $UC_{d.a} \in \{0,1\}^{n^g_d \times T_{d.a}}$ is the day-ahead unit commitment plan, which is decided upon once every $T_{d.a}$ time steps, based on the day-ahead forecasts $\hat{G}_{d.a},\hat{D}_{d.a}$ and current topology $top_t$.
	\item $W_t \in \mathbb{R}_+^{n^g_w}$ is the realized wind generation at time-step $t$. It stays fixed for the actual duration of the time period (1 hour in our simulations).
	\item $D_t \in \mathbb{R}_+^{n^b}$ is the realized load at time-step $t$.
	\item $top_t \in \{0,1\}^{n^l}$ is the network topology at time-step $t$. A line is either up ($1$) or down ($0$), due to two possible events: planned maintenance, or a contingency (line failure), which results in corrective maintenance.
\end{itemize}

\begin{figure}
	\begin{center}
		\includegraphics[scale=0.45]{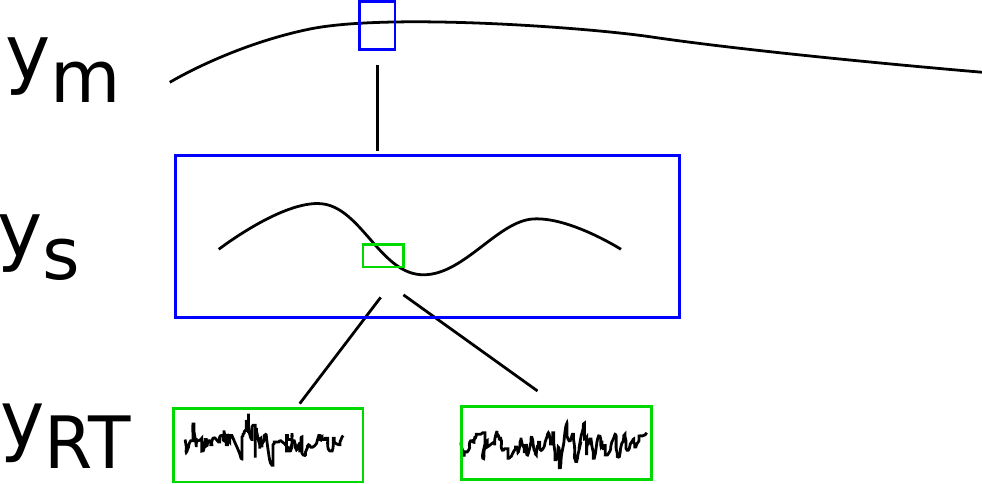}
	\end{center}
	\caption{Hierarchical division of state $\s{t}$ to its three levels of informational states.}
	\label{fig:infomationalStateDivision}
\end{figure}

\subsection{Informational State}
\label{sec:informational_state}
The state-space representation exhibits a division of the state variables to the different temporal evolution processes (time horizons). We formulate this separation of state variables according to the information that is exposed to the decision maker at a decision time point, and refer to it as the \emph{informational state}. Decision makers at different time-horizons are exposed to different amounts of information about the world, hence the higher the time resolution of decisions the more state variables are realized at the time of the decision.  For a general integer $k$, denoting $\s{t}^{1:k}$ to be the sub-vector of $\s{t}$ which includes elements $1$ to $k$, denote $y_m=\s{t}^1=\tau_t$, $y_s=\s{t}^{1:3}=(\tau_t,\hat{W}_{d.a},\hat{D}_{d.a})$ and $y_{RT}=\s{t}$ to be the informational states of the mid-term, short-term and real-time decision resolutions respectively. Fig.~\ref{fig:infomationalStateDivision} presents a separation of the state to informational states. 

A mid-term planner is exposed only to the realization of the $y_m$ part of the state, and can base its mid-term decision $\actionm$ on it and on its expectations of the rest of the state in future times. In the case of $y_s$, on top of being exposed to $\tau_t$, a short-term planner is also exposed to the realization of the day-ahead forecast of generation and load. It is also exposed to the higher-level mid-term decision $\actionm$, however it is not modeled as a part of the informational state. Lastly, as for $y_{RT}$, the real-time planner is exposed to all of the realizations of the components in the state, along with being informed of higher level decisions, i.e., mid-term decision $\actionm$ and short-term decision $\actions$. 

\subsection{Shorter Time-horizon Action-Space} \label{sec:action_space}
Our formulation contains three hierarchical levels of decision making, namely mid-term maintenance, short-term (day-ahead) operational planning, and real-time control. We often refer to the short-term and real-time problems as the \emph{inner problems}. We now present the possible actions in these two inner problems. 
\subsubsection{Short-term Operational Planning Decisions}
The optimal short-term operational planning action 
\begin{equation} \label{eq:optimals}
u_p^* = \argmin_{u_p \in \mathcal{U}_p(u_m)} \quad  C_{p}(y_s,u_m,u_p, u_{RT}^*)
\end{equation} as appears in \eq{eq:optimalS} is defined to be the solution of the unit commitment (UC) schedule. As explained in \S ~\ref{sec:problem_formulation}, the overall cost in \eq{eq:fullMidTermFormulation} is the sum of all real-time interval costs. Therefore the contribution of the solution for this optimization problem is not in directly calculating the overall cost of \eq{eq:fullMidTermFormulation}, rather it is being used as a constraint for the lower-level real-time problem, and as a reference for redispatching costs of real-time operation. 
This UC problem includes DC OPF, with $N-1$ reliability criterion enforced. It also includes wind curtailment and load shedding decisions. This results in a mixed integer-linear program (MILP) that can be solved efficiently using commercial solvers \cite{CPLEX}. The full inner optimization problem is brought in Appendix \ref{sec:unit_commitment}, Problem \ref{eq:unit_commitment}.

Notice that the UC is an optimization program, where the decision is based on the informational state $y_s$ which the decision maker is exposed to when facing a day-ahead planning problem. The informational state $y_s$ contains the wind power and load forecasts $\hat{W}_{d.a},\hat{D}_{d.a}$, which appear in the UC problem formulation. The short-term action-space  $\mathcal{U}_p(u_m)$ in \eq{eq:optimals}, from which the decision variables (as appearing in their detailed form in the full inner optimization problem) are chosen, is the set of possible short-term operational plans. Mid-term decision $\actionm$ dictates which assets are not taking part of the current plan due to maintenance.
\subsubsection{Real-time Control Decisions}
The optimal real-time control action 
\begin{equation} \label{eq:optimalrt}
u_{RT}^* = \argmin_{u_{RT} \in \mathcal{U}_{RT}(u_p)} \quad  C_{RT}(y_{RT},u_p, u_{RT})
\end{equation} as appears in \eq{eq:optimalRT} is defined to be the solution of a DC optimal power flow problem with wind curtailment and load shedding decisions, that follows the original unit commitment plan $\alpha^*$ obtained in the day-ahead planning procedure as detailed in the full inner optimization problem, meaning the participating generators in the power flow at time-step $t$ are those who have $1$'s in their indices in the vector $\alpha^*_t$. It therefore includes re-dispatch decisions, as well as load shedding, wind curtailment and unit re-commitment if necessary, for each time step $t$ individually. It is solved $T_{d.a}$ times sequentially, where each solution at time $t$ is fed to the next one at time $t+1$, e.g., if a contingency takes down a line at time $t$, it is not longer available at time $t+1$, and an $N-1-1$ problem is solved. 
In practice, the real-time optimization problem in \eq{eq:optimalrt} results in a formulation similar to the operational planning formulation in \eq{eq:optimals} that is presented in detail in Appendix \ref{sec:unit_commitment}, therefore we use its formulation and solve it  $T_{d.a}$ times sequentially, with the following adaptations:
\begin{itemize}
	\item it is solved for a single time step $t$, instead of a full day-ahead horizon $T_{d.a}$.
	\item the on/off commitment schedule is no longer a decision variable, rather it is obtained from $u_p^*$ and set as a constraint for each real-time optimization problem at time-step $t$.
	\item wind power and load forecasts $\hat{W}_{d.a},\hat{D}_{d.a}$ for the $T_{d.a}$ time-steps are replaced with their actual realizations $W_t,D_t$.
	\item an additional re-dispatch cost is added to the objective:  $\sum_{i=1}^{n^g_d}\alpha_{t'}^{*,i}|f_P^i(P_{g,t'}^{*,i})-f_P^i(P_{g,t'}^i)|$, assuming re-dispatch cost is the symmetric difference in prices of the generation declared in the day-ahead plan $P_{g,t'}^{*,i}$, and the actual realized power consumed in real-time $P_{g,t'}^i$.
\end{itemize}
Having as an input the full realized state $y_{RT}=\s{t}$ (either by witnessing it in real time, or by sampling future realizations of it), we  solve the real-time control decision problem using the power flow equations and obtain the voltage magnitude and angle at all network nodes. We can then use it to model different related processes, such as aggregated stress effect on equipment failure. However, since this modeling topic requires careful attention and additional research, we currently do not include it.
\subsubsection{Escalation Process}
\begin{figure}
	\begin{center}
		\includegraphics[scale=0.45]{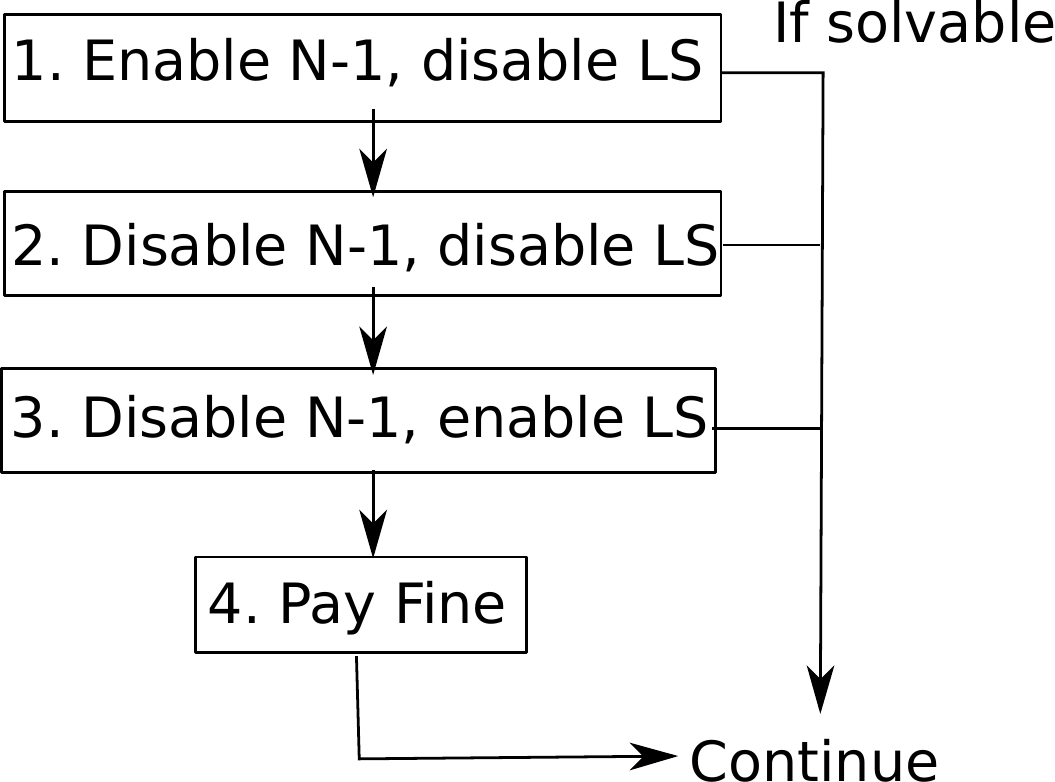}
	\end{center}
	\caption{The short-term and real-time escalation process: for each short-term and real-time inner decision, start by completely removing load-shedding variable $LS$ and its contribution to the objective function from the full inner optimization problem, and attempting to solve it. If feasible solution obtained - finish. If not, remove $N-1$ constraints in the full inner optimization problem, still with load-shedding disabled. If feasible solution obtained - finish. If not, enable load-shedding, still without following $N-1$. If feasible solution obtained - finish. If not, pay large fine (e.g., twice the cost of loss of full network load) and finish.}
	 \label{fig:escalation_process}
\end{figure}

Since high reliability is an important consideration in asset management, and a direct indicator for it is the ability to avoid loss of load, we design a procedure called \emph{escalation process}, which both the optimal short-term and real-time decisions $u_{RT}^*$ and $u_p^*$ follow. It is a four-step escalation procedure, which nurtures two important priorities among TSOs \cite{GARPUR_2.1}: follow $N-1$ criterion as much as possible; avoid load-shedding as much as possible. We therefore introduce this procedure to prioritize usage of  $N-1$ and avoidance of load-shedding. It is brought in detail in Fig. \ref{fig:escalation_process}. 
%operates in the following way:
%\begin{mydescription}{\textbf{Step 1:}}
%	\item[\textbf{Step 1:}]{Completely remove load-shedding variable $LS$ and its contribution to the objective function from problem \ref{eq:unit_commitment}, and attempt to solve it. If feasible solution obtained - finish. If not, continue to step 2.} 
%	\item[\textbf{Step 2:}]{Remove $N-1$ constraints by including only $l=0$ in problem \ref{eq:unit_commitment}, still with load-shedding disabled. If feasible solution obtained - finish. If not, continue to step 3.} 
%	\item[\textbf{Step 3:}]{Enable load-shedding, still without following $N-1$. If feasible solution obtained - finish. If not, continue to step 4.} 
%	\item[\textbf{Step 4:}]{Pay large fine (e.g., twice the cost of loss of full network load) and finish.} 
%\end{mydescription}
In the case of real-time control $u_{RT}$, the procedure starts by following the unit commitment plan as explained before in this section. However, since load shedding is determined as a last resort, in cases where following the day-ahead plan results in load-shedding, the system stops following the day-ahead plan and notes the current time-step $t$ as the beginning of the deviation time. Once deviating from the day-ahead plan, it is no longer enforced in the following time-steps until the end of the day is reached. The system therefore attempts to avoid load shedding by allowing deviation from the day-ahead plan (which consequently incurs re-commitment and re-dispatch costs).

\subsection{Scenario-Space} \label{sec:scenario_space}
The dictionary definition of a scenario is  \emph{``a postulated sequence or development of events''}. We use scenarios as a way to examine plausible future developments in the grid system.  Using a scenario-based approach provides a way of dealing with uncertainties and the complicated interaction between these uncertainties \cite{dembo1991scenario}. 
\subsubsection{Scenario Definition}
\begin{figure}
	\centering
	\includegraphics[scale=0.67]{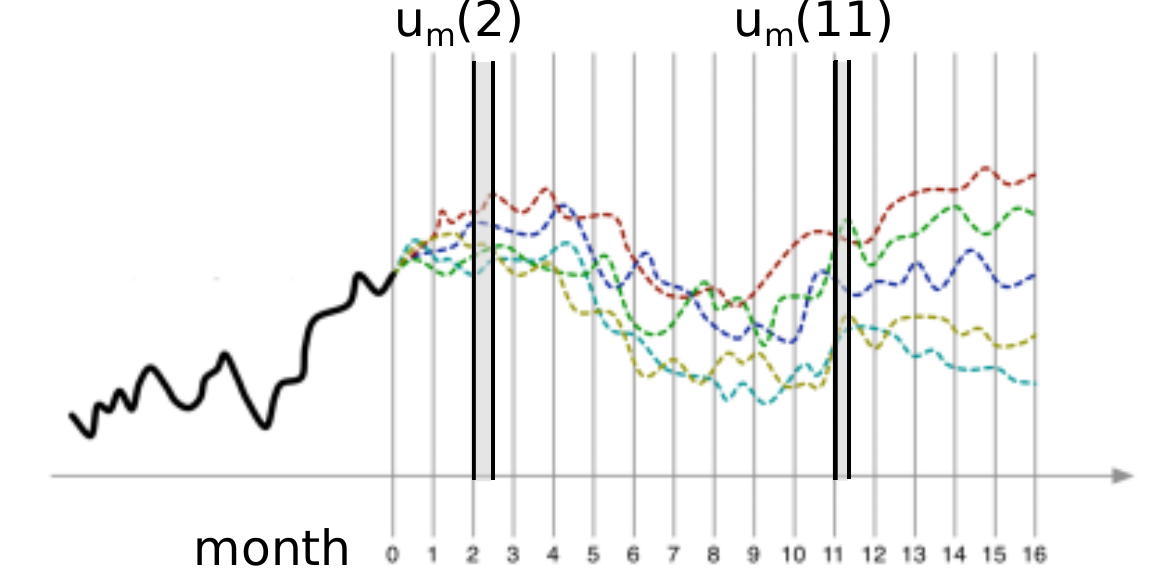}
	%\input{scenarioDescription_v2.pdf_tex}
	
	% Figure \ref{fig:scenarioDescription}b illustrates a real-time decision process of a single mid-term sample path from a zoom in rectangular section in \ref{fig:scenarioDescription}a. The real-time operations are divided to two real-time intervals of the order of 15 minutes. In each interval, the state changes due to operational preventive actions ($u_0$), because of the occurrence of a contingency $\con$, and the evolution of external variables in the state transition.
	\caption{Illustration of sample paths of possible scenarios in the mid-term time horizon. It shows the progression of sampled future scenarios in the mid-term point of view. As time progresses the variance of state $\s{t}$ grows because of increased uncertainty. Each scenario that is generated as a full continuous sequence can be used to test one possible maintenance schedule. A given mid-term planned schedule is shown with maintenance actions in the second and eleventh months, which affect all the mid-term sample paths. }
	\label{fig:scenarioDescription}
\end{figure}

A scenario $\scenario{m} \in {\cal S}^T$ is defined as a sequence of states over the time horizon ${T}$ that are dependent on actions $\actm \in \actionSetm$: \[\scenario{m} = (\s{0},\s{1},\dots,\s{T})\] where $\theta$ is the shorter-term policy, namely the short-term and real-time $N-1$ unit-commitment and DC OPF presented in \S ~\ref{sec:action_space}.
Under the Markovian assumption, possible due to our state and its transition probability definition, the following relation holds:\[ \prob{\scenario{}}=\prob{\s{0}} \cdot \prob{\s{1}|\s{0}}\ldots \prob{\s{T}|\s{T-1}}\] where the state transition probability $\prob{\s{t+1}|\s{t}}$ describes the evolution of the stochastic processes in the system. The  stochasticity stems from the wind power produced in the wind generators $W_t$, the load process $D_t$, and topology of the network $top_t$, as determined by contingency events (unexpected line failure). In Appendix \ref{sec:transition_model} we provide details on the models used for these three probabilistic processes, along with the data and test-cases they are based on.
Figure.~\ref{fig:scenarioDescription} shows an illustrative example of the scenario-space and scenario generation.
% It shows the progression of sampled future scenarios in the mid-term point of view. As time progresses the variance of scenarios grows because of increased uncertainty. Each scenario that is generated as a full continuous sequence can be used to test one possible maintenance schedule.
\subsubsection{Scenario Cost Approximation}
Each scenario has its associated distribution of cost as appears in \eq{eq:scenario_cost}, composed of the summation of real-time atomic costs, due to different events such as redispatching, commitment of generators, and loss of load. To evaluate costs such as these in scenario-based optimization, Monte-Carlo simulation is often used, with two main categories of scenario generation approaches.
 The first is full-trajectory simulation, where all real-time hourly developments are simulated as a full sequence as being done in \cite{jiang2004risk}, while realizing the different uncertainties. For our mid-term problem, which can span over a full year, such an approach will necessitate an intractable number of samples in order to produce a decent evaluation of the scenarios cost, and will incur very high variance of the samples.
 The second category of approaches is based on snapshot sampling of possible static moments of the state of the world and the system. The main problem with such an methodology is the loss of temporal development information, originating in sequential implications of decisions made.
 
Therefore in order to deal with the high complexity of assessing cost/implications of maintenance actions via scenario evaluation, we introduce a novel scenario approximation approach. The \emph{hierarchical window scenario sampling}  is a hybrid version of the two previous sampling methods, which aims at mitigating the disadvantages of each of them. 

\begin{figure} 
	\begin{center}
		\includegraphics[scale=0.32]{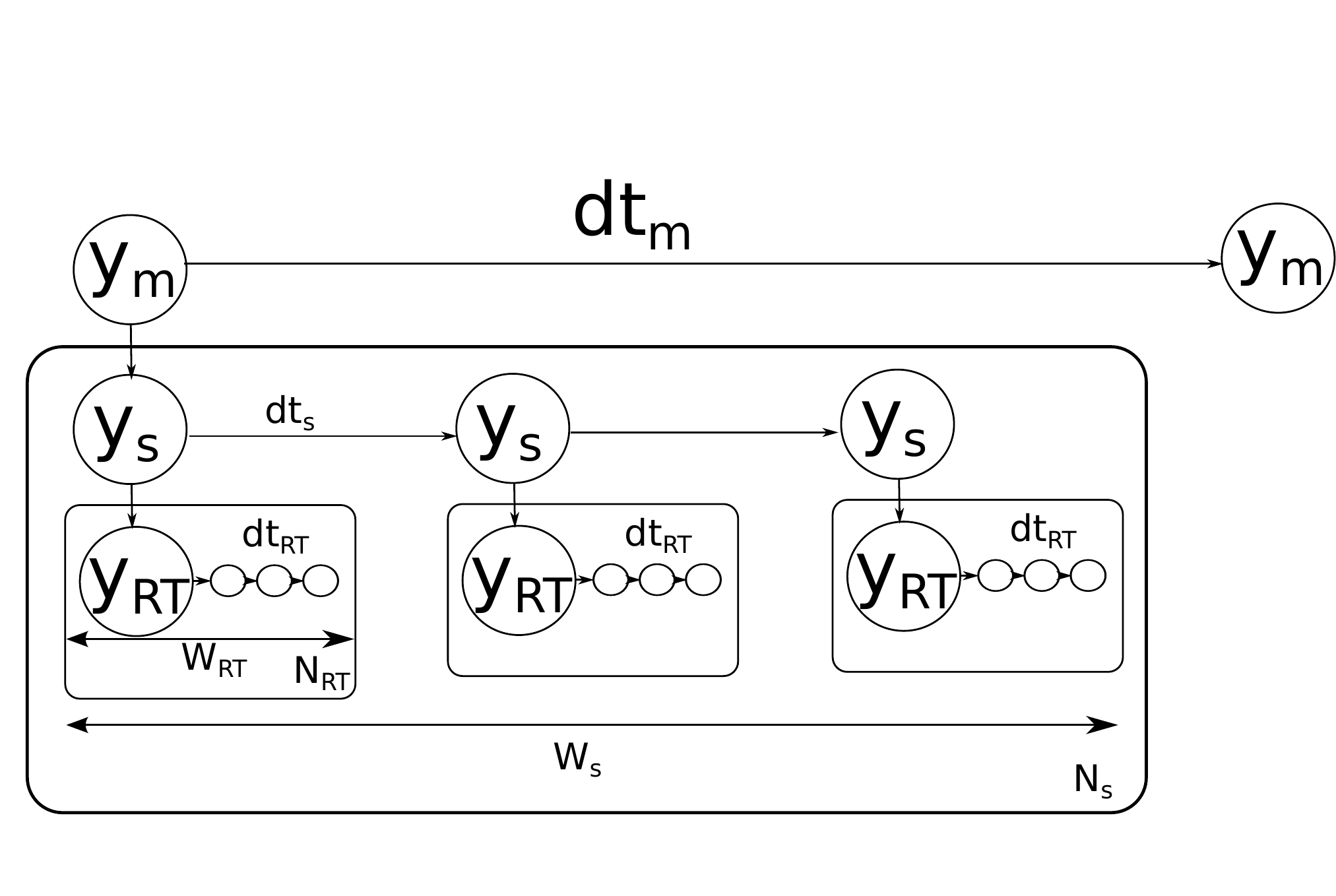}
	\end{center}
	\caption{Hierarchical window scenario sampling approach for scenario cost approximation.}
	\label{fig:scenarioApproximation}
\end{figure}
As visualized in Fig. \ref{fig:scenarioApproximation}, the sampling scheme is done by drawing a sequence of snapshots of the system, for each month (associated with $y_m$), in sequential development of time-ticks $t_m$(=month). Notice that it is governed by the maintenance schedule $\actionm$. Then, to approximate the cost of each month we draw $N_s$ samples of $W_s$ sequential days (trajectories of $y_s$ at time resolution $t_s$=day). Notice that the short-term inner optimization problem is solved for each day in each trajectory. Next, for each day ($y_{s}$) we simulate the real-time progression by sampling $N_{RT}$ trajectories of length $W_{RT}$ (=$24$) of the hourly sequential $y_{RT}$\footnote{The choice of the window lengths $W_s$ and $W_{RT}$ controls the level of interpolation between the completely sequential scenario sampling of all $T$ time steps, and the alternative completely static approach of solely sampling snapshots of states, with no temporal relation between them. Essentially, they arbitrate between the bias and variance of the sampling process. Full trajectory sampling has low bias but high variance, while static snapshot sampling lowers the variance, though it introduces bias due to its simplicity and choice of times of snapshots.}. For each $y_{RT}$ we calculate the real-time cost by solving the real-time OPF problem with its associated sampled (realized) uncertainties.
%The bottom part of the figure shows a zoom in section of a single scenario in real-time point of view.
%Real-time intervals are in the order of 15 minutes. In the real-time interval is state is modeled as a transition process between certain events. For example, first part of the state transition due to an operational preventive action taken by the operator $u_0$ based on the knowledge of possible contingencies $\curlyN$, then a stochastic transitions can occur due to a contingency $\con$, and finally,  state variables can evolve before the next RT interval. Transitions in state can effect separate parts of the state vector. Variable that change on a slower time horizon than the current point of view are assumed to be static during the interval. Differently than actions in the mid-term, here the actions are selected differently for each scenario and are state dependent.
\section{Distributed Simulation-Based Cross Entropy Optimization Algorithm} \label{sec:cross_entropy}
Problem \ref{eq:fullMidTermFormulation} is complex, being a non-convex combinatorial stochastic optimization problem, with hierarchical inner lower-level mixed integer-linear programs of shorter-time optimization problems. 
Since no explicit analytical form of the objective and constraints can be obtained for this problem, in order to asses the cost of scenario from \eq{eq:scenario_cost} while obtaining its required inner-decisions from \eqs{eq:optimalS}{eq:optimalRT}, we choose a simulation-based optimization approach. It is performed with distributed Monte-Carlo sampling, where multiple solutions in $\actionSetm$ are being assessed in parallel on multiple servers. Each month of such solution assessment is itself assessed in parallel. In turn it solves inner MILP programs of short-term decisions and their consecutive LP programs of real-time decision.

For the optimization algorithm in Problem (\ref{eq:fullMidTermFormulation}) we use the cross-entropy (CE) optimization algorithm \cite{de2005tutorial}. The CE algorithm has shown to be useful in other works for solving power system combinatorial problems \cite{ernst2007cross}. Briefly, CE is a randomized percentile optimization method for solving difficult combinatorial programs. It iteratively performs a step of generating random data sample according to some parametric distribution, followed by a step of updating the parameters of this distribution based on the data to produce a ''better'' sample in the next iteration.
In our case, this distribution is of the assessed solutions $\actionm$, where at each iteration of the cross-entropy algorithm, the top percentile of the solutions assessed is used for updating their distribution. 
%This methodology is presented graphically in Fig. \ref{fig:methodology}.
\section{Experimental Results} \label{sec:experiment_results}
We run our experiments on a Sun cluster with Intel(R) Xeon(R) cpus  $@2.53$GHz, containing 300 cores, each with 2GB of memory. 
All code is written in matlab 
%\cite{matlab}
. We use YALMIP \cite{YALMIP} to model the full inner optimization problem both in its short-term and real-time versions. It is then solved using CPLEX \cite{CPLEX}.

As a test-case we use a 5-bus modified system as appears in MATPOWER simulation tool \cite{zimmerman2011matpower}, with 5 dispathable generators, based on PJM 5-bus system \cite{li2010small}. We adopt updated generator parameters from Kirschen et. al \cite{pandzic2013comparison}, namely their capacities, min-output, ramp up/down limits, min up/down times, price curve and start-up costs. Specifically, these are unit types $1,~3,~5,~7,~9$ as appear in \cite{pandzic2013comparison} that replace the original ones in the 5-bus system in that specified order. They are composed of $1$ open cycle gas turbine, $3$ combined cycle gas turbines, and $1$ nuclear plant, which are more representative of current power production technology. We also include $2$ non-dispatchable wind generators in buses $1,~2$, with capacities and daily generation profiles, based on real historical records from the US as published in \cite{UW_website}. Peak loads and daily demand profile are based on real data, taken from \cite{UW_website}, and are rescaled to fit the 5-bus network. For more information on the wind generation and load distributions used in our experiments please refer to Appendix \ref{sec:transition_model}. Value of lost load cost is set to $VOLL=1000[\frac{\$}{MWh}]$, taken from \cite{dvorkin2015hybrid} and wind-curtailment price is set to $C_{WC}=100[\frac{\$}{MWh}]$, taken from \cite{loisel2010valuation}. Transmission line maintenance cost is set to $C_{\actionm}=5000\$$ for all assets, taken from \cite{costing2012}. We solve Problem (\ref{eq:fullMidTermFormulation}) for a time-horizon of $T_M=8$ months. Each cross-entropy iteration assesses the cost of $75$ possible maintenance schedules in parallel, where each evaluated month of such a schedule is split to a different server, and is simulated as $W_s=3$ consecutive days, resulting in $3$ unit-commitment solutions of the full inner optimization problem for each such trajectory. For each of those days, we simulate $N_{RT}=30$ samples of real-time trajectories of $W_{RT}=24$ hours per each sample. Each such trajectory is a realization of the actual wind power $W_t$, load $D_t$, and contingencies $top_t$ that occurred in that duration, using which the real-time control decision in \eq{eq:optimalrt} is calculated and its cost is obtained. The maintenance feasibility constraint in \eq{eq:midTermFeasibility} is set to allow no more than maintenance of a single asset per month, and no more than a single maintenance per each asset throughout the planning horizon.
In order to be able to compare our solution to maintenance routines TSOs are likely to use, we run our scenario cost assessment on three intuitive heuristic maintenance plans: an 'oldest-first' maintenance plan, where at each month the least recently maintained asset in the network is chosen for maintenance; an 'age-threshold' maintenance plan, where assets are maintained when reaching a certain effective age threshold; a 'cyclic' maintenance plan, where assets are maintained sequentially, in a cyclic manner.

\begin{figure} 
	\begin{center}
	\includegraphics[scale=0.3]{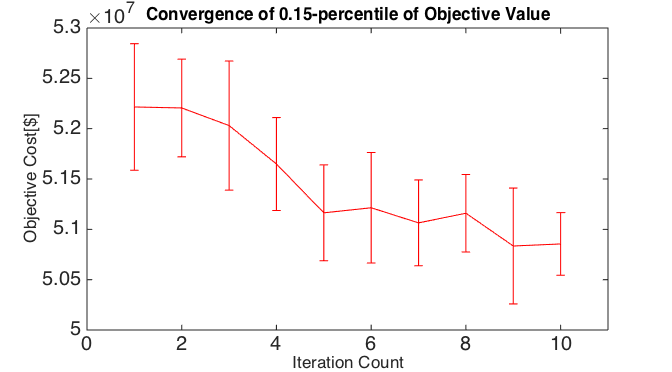}
	\end{center}
	\caption{Convergence of the mean $0.15$-percentile cost of the optimization algorithm. Error bars show the standard deviation.}
	\label{fig:convergence}
\end{figure}

Fig.~\ref{fig:convergence} presents the convergence of our algorithm. The figure shows the mean and standard deviation of the  top-$0.15$-percentile of the cost (as appears in \eq{eq:midTermObj}) of all maintenance solutions $\actionm \in \actionSetm$ assessed at each CE iteration.
Fast convergence is achieved after only $10$ iterations, as can be expected from the CE algorithm \cite{ernst2007cross}. As iterations progress, the mean cost of the sampled scenarios decreases, as well as their standard deviation. This is due to the convergence of the distribution parameters in the CE algorithm, where samples become more concentrated in local optima of the solution-space $\actionSetm$. An interesting phenomenon can be witnessed in iterations $9$ and $10$, where  the mean stayed the same while the standard deviation dropped, due to natural preservation and duplication of the best solutions, and the increased number of their samples.
%The standard deviation of the first iteration ... 

\begin{figure} 
	\includegraphics[scale=0.3]{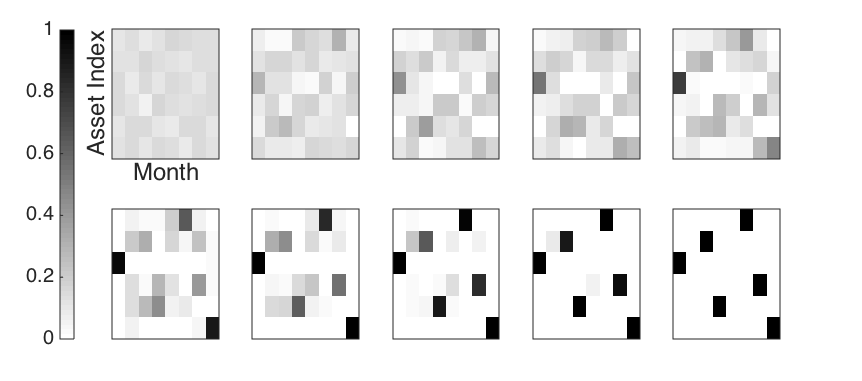} 
	\caption{Visual representation of the frequencies of the maintenance matrices along iterations, showing convergence to a single, best schedule.}
	\label{fig:planimage}
\end{figure}

Fig.~\ref{fig:planimage} visually presents the different maintenance schedules drawn for assessment across iterations. Each entry of the matrices is colored according to the portion of assessed solutions which had $1$'s in that entry, where 'white' means none of the entries were chosen to be '$1$', and 'black' means all solutions of that iteration drew a '1' in that entry. 
\begin{figure} 
	\begin{center}
		\includegraphics[scale=0.23]{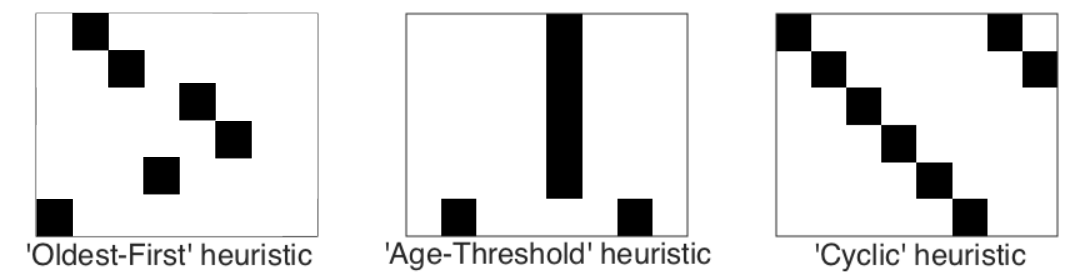}
	\end{center}
	\caption{Visual representation of the three alternative heuristic maintenance strategies.}
	\label{fig:heuristics}
\end{figure}The quick convergence of maintenance schedules to the (local) optimal solution is visualized by the gray colors which quickly turn either white or black.
In addition, Fig.~\ref{fig:heuristics} uses the same visual concept to present the three maintenance heuristics we compare our solution to.  
%Notice how the two last matrices (the optimization solution and the 'oldest-first') are almost similar, however asset $\#3$ is maintained first in the optimization solution due to its criticality in the system.

\begin{figure} 
	\begin{center}
	\includegraphics[scale=0.25]{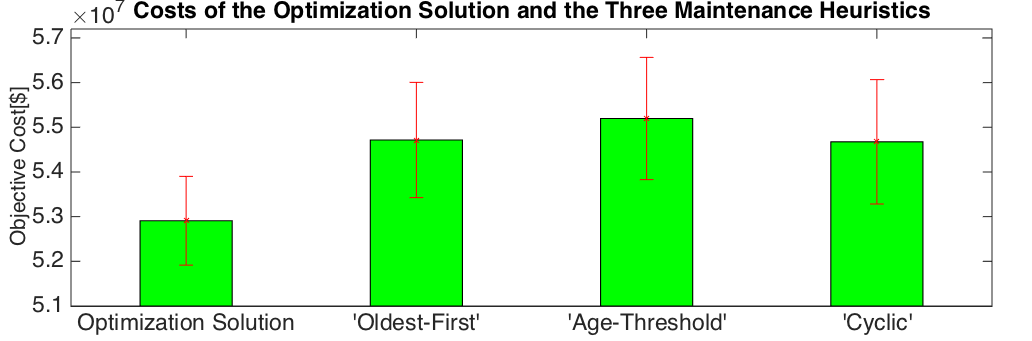} 
	\end{center}
	\caption{Costs and standard deviations of the optimization solution and the three alternative heuristic maintenance strategies.}
	\label{fig:results}
\end{figure}

Lastly, in Fig.~\ref{fig:results} we bring the mean and standard deviation of the costs of the best solution found by our optimization algorithm and the three different maintenance heuristics, evaluated using $50$ samples. Our solution is significantly cheaper than all three alternative heuristics. The costs' $1$-standard deviation intervals are almost not in conjunction, suggesting high confidence in the statistical significance. 
%[!ht]
%\begin{table}
%	%% increase table row spacing, adjust to taste
%	\renewcommand{\arraystretch}{1.2}
%	%% if using array.sty, it might be a good idea to tweak the value of
%	%% \extrarowheight as needed to properly center the text within the cells
%	%
%	\caption{Maintenance schedule mean and standard-deviation of objecrive cost}
%	\label{tab:reults}
%	\noindent
%	\centering
%	\begin{minipage}{\linewidth} %Use the minipage environment to footnote tables
%		\renewcommand\footnoterule{\vspace*{-5pt}} %to remove the horizontal rule above the table footnote
%		\begin{center}
%			\begin{tabular}{ |l | c | c | c |}
%				\hline 
%				Maintenance Schedule & Mean cost [\$] & std of cost [\$] ]\\ \hline \hline
%				Solution  of Problem (\ref{eq:fullMidTermFormulation})   & $2.5\cdot 10^6$ & $2.5\cdot 10^4$ \\ \hline
%				'Oldest-first' heuristic   & $2.75\cdot 10^6$ & $3\cdot 10^4$ \\ \hline 
%%				\Xhline{2\arrayrulewidth}
%%				\textbf{Tree Search, H=1} & \textbf{512,850} & \textbf{2.5} & last\\ \hline
%%				Tree Search, H=3 & 512,217 & 240 & last\\ \hline
%%				Sub-sampled Tree Search, H=3 & 512,850 & 85& last \\ \hline
%%				Back Sweep & 511,500 & 60& last \\ \hline
%			\end{tabular}
%		\end{center}
%	\end{minipage}
%\end{table}
%   Please see CASEFORMAT for details on the case file format.
%
%   Based on data from ...
%     F.Li and R.Bo, "Small Test Systems for Power System Economic Studies",
%     Proceedings of the 2010 IEEE Power & Energy Society General Meeting

\section{Conclusion} \label{sec:conclusion}
The asset management problem requires careful attention when planning a maintenance schedule, due to the hierarchical structure of this problem, where several layers of decision making need to be accounted for. This makes the asset management task a very challenging problem for mid-term planners within a TSO.
The scenario assessment framework developed in this work enables to evaluate the multiple complex implicit implications a maintenance schedule inflicts on a power system. We harness the power of distributed computing for the evaluation of these implications and for an optimization algorithm that seeks optimal maintenance plans, which a human expert can possibly not consider, as can be seen from our comparison to several simple-yet-convincing heuristic maintenance strategies. Our framework can also be used for assessing the implications of a predefined maintenance schedule considered by an expert.

In this work we use simplifying assumptions, such as that maintenance resets assets' effective age, the effective age dictates the failure probability of an asset, and that wind is distributed according to a multivariate normal random variable. These assumptions might render too simplistic for real-life power systems, however they allow for good scalability of the method and form an initial starting point. 

The method presented here will not scale well to realistic grids, with thousands of nodes, generators and loads. To deal with the high dimensional combinatorial structure of the maintenance schedule and the inner decisions such as unit commitment, approximated 'proxy' methods are necessary. A recent example for such can be found in \cite{dalal2016hierarchical}.
These functions will allow fast assessment calculation for a single schedule,  provide a generalization mechanism over possible future shorter time decisions, and enable more realistic assumptions. 

In continuation to this work, in addition to using such proxy modules, we plan on adding to our formulation chance-constraints for ensuring high reliability; considering more than binary actions for more realistic maintenance; using larger test-cases such as IEEE RTS-96.
\bibliographystyle{IEEEtran}
\bibliography{CandidacyExam}

\begin{thebibliography}{xx}

\harvarditem[Defourny et~al.]{Defourny, Ernst and Wehenkel}{2012}{defourny2012}
Defourny, B., Ernst, D. and Wehenkel, L.: 2012, Scenario trees and policy
  selection for multistage stochastic programming using machine learning, {\em
  INFORMS Journal on Computing} .

\harvarditem[Dupacova et~al.]{Dupacova, Consigli and Wallace}{2000}{dupacova}
Dupacova, J., Consigli, G. and Wallace, S.: 2000, Scenarios for multistage
  stochastic programs, {\em Annals of Operations Research} {\bf
  100}(1-4),~25--53.

\harvarditem{{GARPUR Consortium}}{2014}{D2.1}
{GARPUR Consortium}: 2014, Functional analysis of reliability management, 7th
  framework programme, EU Commission grant agreement 608540.
\newline\harvardurl{https://project.sintef.no/eRoom/energy2/GARPUR-FP7proposal/0_1489e}

\harvarditem{{GARPUR Consortium}}{2015}{D6.1}
{GARPUR Consortium}: 2015, Functional workflow of short-term \& real-time
  decision making processes, 7th framework programme, EU Commission grant
  agreement 608540.
\newline\harvardurl{https://project.sintef.no/eRoom/energy2/GARPUR-FP7proposal/0_a6f6}

\harvarditem[Growe-Kuska et~al.]{Growe-Kuska, Heitsch and
  Romisch}{2003}{growe-kuska}
Growe-Kuska, N., Heitsch, H. and Romisch, W.: 2003, Scenario reduction and
  scenario tree construction for power management problems, {\em Power Tech
  Conference Proceedings, 2003 IEEE Bologna}, Vol.~3, p.~7.

\harvarditem{Rockafellar and Wets}{1991}{Rockafellar}
Rockafellar, R.~T. and Wets, R. J.-B.: 1991, Scenarios and policy aggregation
  in optimization under uncertainty, {\em Math. Oper. Res.} {\bf
  16}(1),~119--147.

\end{thebibliography}


\begin{thebibliography}{10}
\providecommand{\url}[1]{#1}
\csname url@samestyle\endcsname
\providecommand{\newblock}{\relax}
\providecommand{\bibinfo}[2]{#2}
\providecommand{\BIBentrySTDinterwordspacing}{\spaceskip=0pt\relax}
\providecommand{\BIBentryALTinterwordstretchfactor}{4}
\providecommand{\BIBentryALTinterwordspacing}{\spaceskip=\fontdimen2\font plus
\BIBentryALTinterwordstretchfactor\fontdimen3\font minus
  \fontdimen4\font\relax}
\providecommand{\BIBforeignlanguage}[2]{{%
\expandafter\ifx\csname l@#1\endcsname\relax
\typeout{** WARNING: IEEEtran.bst: No hyphenation pattern has been}%
\typeout{** loaded for the language `#1'. Using the pattern for}%
\typeout{** the default language instead.}%
\else
\language=\csname l@#1\endcsname
\fi
#2}}
\providecommand{\BIBdecl}{\relax}
\BIBdecl

\bibitem{karangelos2015probabilistic}
\BIBentryALTinterwordspacing
E.~Karangelos and L.~Wehenkel, ``Probabilistic reliability management approach
  and criteria for power system real-time operation,'' \emph{. To appear In
  Proc. of Power Systems Computation Conference (PSCC 2016), 2016}. [Online].
  Available: \url{http://hdl.handle.net/2268/193403}
\BIBentrySTDinterwordspacing

\bibitem{GARPUR_2.1}
{GARPUR Consortium}, ``D2.1, functional analysis of reliability management,''
  [Online]. Available: http://www.garpur-project.eu/deliverables, Oct. 2014.

\bibitem{GARPUR_5.1}
{GARPUR {C}onsortium}, ``D5.1, functional analysis of asset management
  processes,'' [Online]. Available: http://www.garpur-project.eu/deliverables,
  Mar. 2015.

\bibitem{abiri2013two}
A.~Abiri-Jahromi, M.~Parvania, F.~Bouffard, and M.~Fotuhi-Firuzabad, ``A
  two-stage framework for power transformer asset maintenance management --part
  i: Models and formulations,'' \emph{Power Systems, IEEE Transactions on},
  vol.~28, no.~2, pp. 1395--1403, 2013.

\bibitem{jiang2004risk}
Y.~Jiang, Z.~Zhong, J.~McCalley, and T.~Voorhis, ``Risk-based maintenance
  optimization for transmission equipment,'' in \emph{Proc. of 12th Annual
  Substations Equipment Diagnostics Conference}, 2004.

\bibitem{entso-e}
ENTSO-E, \emph{ENTSO-E guideline for cost-benefit analysis for grid development
  projects}, 2013.

\bibitem{fu2007security}
Y.~Fu, M.~Shahidehpour, and Z.~Li, ``Security-constrained optimal coordination
  of generation and transmission maintenance outage scheduling,'' \emph{Power
  Systems, IEEE Transactions on}, vol.~22, no.~3, pp. 1302--1313, 2007.

\bibitem{puterman1990markov}
M.~L. Puterman, ``Markov decision processes,'' \emph{Handbooks in operations
  research and management science}, vol.~2, pp. 331--434, 1990.

\bibitem{CPLEX}
``{IBM ILOG CPLEX Optimizer},''
  http://www-01.ibm.com/software/integration/optimization/cplex-optimizer/,
  2010.

\bibitem{dembo1991scenario}
R.~S. Dembo, ``Scenario optimization,'' \emph{Annals of Operations Research},
  vol.~30, no.~1, pp. 63--80, 1991.

\bibitem{de2005tutorial}
P.-T. De~Boer, D.~P. Kroese, S.~Mannor, and R.~Y. Rubinstein, ``A tutorial on
  the cross-entropy method,'' \emph{Annals of operations research}, vol. 134,
  no.~1, pp. 19--67, 2005.

\bibitem{ernst2007cross}
D.~Ernst, M.~Glavic, G.-B. Stan, S.~Mannor, and L.~Wehenkel, ``The
  cross-entropy method for power system combinatorial optimization problems,''
  in \emph{2007 Power Tech}, 2007.

\bibitem{YALMIP}
\BIBentryALTinterwordspacing
J.~Löfberg, ``Yalmip : A toolbox for modeling and optimization in {MATLAB},''
  in \emph{Proceedings of the CACSD Conference}, Taipei, Taiwan, 2004.
  [Online]. Available: \url{http://users.isy.liu.se/johanl/yalmip}
\BIBentrySTDinterwordspacing

\bibitem{zimmerman2011matpower}
R.~D. Zimmerman, C.~E. Murillo-S{\'a}nchez, and R.~J. Thomas, ``Matpower:
  Steady-state operations, planning, and analysis tools for power systems
  research and education,'' \emph{Power Systems, IEEE Transactions on},
  vol.~26, no.~1, pp. 12--19, 2011.

\bibitem{li2010small}
F.~Li and R.~Bo, ``Small test systems for power system economic studies,'' in
  \emph{Power and Energy Society General Meeting, 2010 IEEE}.\hskip 1em plus
  0.5em minus 0.4em\relax IEEE, 2010, pp. 1--4.

\bibitem{pandzic2013comparison}
H.~Pandzic, T.~Qiu, and D.~S. Kirschen, ``Comparison of state-of-the-art
  transmission constrained unit commitment formulations,'' in \emph{Power and
  Energy Society General Meeting (PES), 2013 IEEE}.\hskip 1em plus 0.5em minus
  0.4em\relax IEEE, 2013, pp. 1--5.

\bibitem{UW_website}
``Renwable energy analysis lab, electrical engineering department, university
  of washington,''
  \url{http://www.ee.washington.edu/research/real/library.html}, note =
  {Accessed: 2015-09-16}.

\bibitem{dvorkin2015hybrid}
Y.~Dvorkin, H.~Pandzic, M.~Ortega-Vazquez, D.~S. Kirschen \emph{et~al.}, ``A
  hybrid stochastic/interval approach to transmission-constrained unit
  commitment,'' \emph{Power Systems, IEEE Transactions on}, vol.~30, no.~2, pp.
  621--631, 2015.

\bibitem{loisel2010valuation}
R.~Loisel, A.~Mercier, C.~Gatzen, N.~Elms, and H.~Petric, ``Valuation framework
  for large scale electricity storage in a case with wind curtailment,''
  \emph{Energy Policy}, vol.~38, no.~11, pp. 7323--7337, 2010.

\bibitem{costing2012}
P.~Brinckerhoff, ``Electricity transmission costing study,'' 2012.

\bibitem{dalal2016hierarchical}
G.~Dalal, E.~Gilboa, and S.~Mannor, ``Hierarchical decision making in
  electricity grid management,'' \emph{arXiv preprint arXiv:1603.01840}, 2016.

\bibitem{amaniampong1996monte}
G.~Amaniampong and C.~Burgoyne, ``Monte-carlo simulations of the time dependent
  failure of bundles of parallel fibres,'' \emph{European Journal of Mechanics,
  A/Solids}, vol.~15, no.~2, pp. 243--266, 1996.

\end{thebibliography}
\newpage
%\clearpage
\appendices
%\gdef\thesection{Appendix \Alph{section}}
%\addtocontents{toc}{\protect\contentsline{chapter}{Appendix:}{}}
%\begin{appendix}

	\section{Short-term Unit Commitment and Real-time OPF actions} \label{sec:unit_commitment}
In the case of DC power flow, voltage magnitudes and reactive powers are eliminated from the problem and real power flows are modeled as linear functions of the voltage angles. This results in a mixed integer-linear program (MILP) that can be solved efficiently using commercial solvers \cite{CPLEX}. The exact unit-commitment problem formulation is the following:
%	& u_p^* = \argmin_{u_p \in \mathcal{U}_p(u_m, y_s)} \quad  C_{p}(u_m,u_p, u_{RT}^*) \\= \argmin_{\alpha,\Theta,P_g,t,WC,LS} &\quad  &\sum_{t'=t}^{t+T{d.a}} \left[\sum_{i=1}^{n^g_d}\left(\alpha_{t'}^if_P^i(P_{g,t'}^i) + \alpha_{t'}^i(1-\alpha_{t'-1}^i)SU_i(t_{\text{off}}^i) \right)\right.  \label{eq:objective}\\
\begin{subequations}  \label{eq:unit_commitment}
	\begin{align}
	& u_p^* = \argmin_{u_p \in \mathcal{U}_p(u_m, y_s)}   C_{p}(u_m,u_p, u_{RT}^*) = \argmin_{\alpha,\Theta,P_g,t,WC,LS} \nonumber \\
	& \sum_{t'=t}^{t+T{d.a}} \left[\sum_{i=1}^{n^g_d}\left(\alpha_{t'}^if_P^i(P_{g,t'}^i) + \alpha_{t'}^i(1-\alpha_{t'-1}^i)SU_i(t_{\text{off}}^i) \right)\right.   \nonumber \\
	&\quad \quad \quad + \left. \sum_{iw=1}^{n^g_w}WC_{t'}^{iw}\cdot C_{WC} + \sum_{ib=1}^{n^b}LS_{t'}^{ib}\cdot VOLL \right] \label{eq:objective}\\
	&\text{subject\ to}  \\
	& g_{P,t'}^l(\Theta^l,\alpha,P_g)=~B_{\text{bus}}^l \Theta_{t'}^l + P_{BUS,\text{shift}}^l + \hat{D}_{d.a,t'} \label{eq:power_balance} \\
	& ~~+G_{sh} - LS_{t'}  - (\hat{W}_{d.a,t'}-WC_{t'})  - C_g (\alpha_{t'}.* P_g) = 0 \nonumber\\
	& h_{f,t'}^l(\Theta_{t'}^l) =  B_f^l \Theta_{t'}^l + P_{f,\text{shift}}^l - F_{max}^l \leq 0 \\
	& h_{t,t'}^l(\Theta_{t'}^l) = B_f^l \Theta_{t'}^l - P_{f,\text{shift}}^l - F_{max}^l \leq 0 \label{eq:to_line_limits} \\
	& \theta_i^{\text{ref}} \leq \theta_{i,t'}^l \leq \theta_i^{\text{ref}},  \quad i \in {\cal I}_{\text{ref}} \label{eq:angle_limits}\\
	& \alpha_{t'}^ip_g^{i,\text{min}} \leq p_{g,t'}^i \leq \alpha_{t'}^ip_g^{i,\text{max}}, \quad  i=1,\dots,n^g_d\\
	& 0 \leq WC_{t'}^{iw} \leq \hat{W}_{d.a,t'}^{iw}, \quad  iw=1,\dots,n^g_w\\
	& 0 \leq LS_{t'}^{ib} \leq \hat{D}_{d.a,t'}^{ib}, \quad  ib=1,\dots,n^b \label{eq:LS_limit}\\
	& t_{\text{off}}^i \geq  t_{\text{down}}^i, \quad  i=1,\dots,n^g_d \label{eq:min_down}\\
	& t_{\text{on}}^i \geq  t_{\text{up}}^i,\quad   i=1,\dots,n^g_d \label{eq:min_up}\\
	&l=0,1,\dots,n^l_t \\
	&t'=t,\dots,t+T{d.a}
	%		&t'=t,t+1,\dots,t+T{d.a}
	\end{align} 
\end{subequations}

where
\begin{itemize}
	\item $l$ is the index of a line that is offline. $l=0$ means that all lines are connected and online. (lines that are under maintenance are not counted in $n^l_t$ to begin with).
	\item $\alpha \in \{0,1\}^{n^g_d \times T{d.a}}$ is the commitment (on/off) status of all dispatchable generators, at all time-steps.
	\item $\Theta \in [-\pi,\pi]^{ n^b \times (n^l+1) \times T{d.a}}$ are the different voltage angle vectors for the different network layouts, for all time steps.
	%		\item $P_g \in \mathbb{R}_+^{n^g_d}$ are the power outputs of the dispatchable generators.
	\item $P_g \in \mathbb{R}_+^{n^g_d \times T{d.a}}, WC \in \mathbb{R}_+^{n^g_w \times T{d.a}}, LS \in \mathbb{R}_+^{n^b \times T{d.a}}$ are the dispatchable generation, wind curtailment and load shedding decision vectors, with $f_P, C_{WC}, VOLL$ as their corresponding prices.
	\item $t_{\text{down}}^i, t_{\text{up}}^i$ are the minimal up and down times for generator $i$, after it had been off/on for $t_{\text{off}}^i$/$t_{\text{on}}^i$.
	\item $SU_i(t_{\text{off}}^i)$ is the start-up cost of dispatchable generator $i$ after it had been off for $t_{\text{off}}^i$ time-steps.
	\item $g_{P,t'}^l(\Theta^l,\alpha,P_g)$ is the overall power balance equation for line $l$ being offline.
	\item $B_{\text{bus}},P_{BUS,\text{shift}}$ are the nodal real power injection linear relation terms.
	\item $B_f,P_{f,\text{shift}}$ are the linear relation terms of the branch flows at the \emph{from} ends of each branch (which are the minus of the \emph{to} ends, due to the lossless assumption).
	\item $G_{sh}$ is the vector of real power consumed by shunt elements.
	\item $C_g$ is the generator-to-bus connection matrix, $(\alpha_{t'}.* P_g)$ is the dot-product of the two vectors.
	\item $F_{max}$ are the line flow limits.
	\item ${\cal I}_{\text{ref}}$ is the set of indices of reference buses, with $\theta_i^{\text{ref}}$ being the reference voltage angle.
	\item $p_g^{i,\text{min}},p_g^{i,\text{max}}$ are the minimal and maximal power outputs of generator $i$.
\end{itemize}
More information on the DC approximation can be found in \cite{zimmerman2011matpower}.\\
Constraints in \eqs{eq:power_balance}{eq:to_line_limits} ensure load balance and network topology constraints.\\
Constraints in \eqs{eq:angle_limits}{eq:LS_limit} restrict the decision variables to stay within boundary, namely voltage angle limits, generator minimal and maximal power output range, wind curtailment and load shedding limits. \\
Constraints in \eqs{eq:min_down}{eq:min_up} bind the different time steps to follow generator minimal up and down time thermal limits.\\

\section{Distributions of the Stochastic Processes within the Model}  \label{sec:transition_model}
The stochastic processes in the system are the wind power produced in the wind generators $W_t$, the load process $D_t$, and topology of the network $top_t$, as determined by contingency events (unexpected line failure). In this section we provide details on the models used for these three probabilistic processes, along with the data and test-cases they are based on.
\subsubsection{Wind Power Distribution}
The wind generation capacities for buses with wind generators attached are taken from \cite{UW_website}, along with their daily mean profile. 
%In addition, an annual wind profile is adopted from \cite{seasonalWindProfile}. 
The wind process mean $\mu_w(t)$ is therefore obtained by the formula \[\mu_w(t) = \mu_w(t_D)\cdot p_{w,\text{annual}}(t_M)\], where $\mu_w(t_D) \in \mathbb{R}_+^{n^g_w}$ is the daily wind mean profile at time-of-day $t_D$, and $p_{w,\text{annual}}(t_M) \in [0,1]$ is the annual wind profile relative to its peak at month $t_M$ of the year. \\
Wind generation process $W_t$ is a multivariate, normally distributed random variable \[W_t \sim \mathcal{N}\left(\mu_w(t),diag((p_{w,\sigma}\cdot \mu_w(t))^2)\right)\] where $p_{w,\sigma} \in [0,1]$ is a constant ($0.15$ in most simulations) that multiplies the mean $\mu_w(t)$, to obtain a standard deviation that is a fixed fraction of the mean. $diag(x)$ is a square diagonal matrix, with the elements of $x$ as its diagonal, so different wind generators are assumed uncorrelated.  $W_t$ is truncated to stay in the range between $0$ and the generator's capacity.
\subsubsection{Load Distribution}
Load $D_t$ is assumed to follow the same normal distribution as the wind, with the same formula containing peak loads and daily profiles for each bus $\mu_d(t_D) \in \mathbb{R}_+^{n^b}$ with values taken from \cite{UW_website}.
% and an annual load profile $p_{d,\text{annual}}(t_M)$ adopted from \cite{seasonalDemandProfile}. 
 Fraction of mean for standard deviation is set to be $p_{d,\sigma}=0.02$.
\subsubsection{Contingency Probability}
To calculate the failure probability of the lines, we use a life model based on the Weibull probability distribution \cite{amaniampong1996monte}. 
The probability distribution function for an asset life time with stress history $\sigma(t)$ at time $t>0$ is
\[ H(t;\sigma) = 1-\exp \left( -\Psi \left[ \int_0^t k(\sigma(t')) dt'\right] \right) \]
where $k(\cdot)$ represents the rate at which the stress history effects the probability of failure, and $\Psi$ is the lifetime model distribution. In our model, we use an exponential-law break down rule $k(\sigma) = \alpha \exp(\gamma \sigma)$, and a Weibull distribution to model the lifetime $\Psi(x) = \nu x^s,~x>0$. We also simplify the model by assuming constant stress. In addition, since in our case we assume a 'reseting' effect of a maintenance action,  the contingency probability only depends on the effective age $\tau$, resulting in 
\[ H(\tau) = 1-\exp \left( -\nu \left[  \alpha \exp(\gamma \tau) \right]^s \right) \]
, where $\nu,\alpha,\gamma,s$ are the model parameters. 
%Mathematical models are needed to asses the likelihood of an asset failure and to represent the effect of maintenance. An asset failure can be caused due to component strength weakening due to age, random external events such as a adverse weather, or a combination of the two, see Fig.~\ref{fig:cigreAsset} and section 2.3.4 in \cite{cigrec116}. For example, it could be assumed that an aging line is more susceptible to random stress such as wind damage, than a newly constructed line. There are a variety of asset life mathematical models in the literature. Mathematical models may be deterministic or probabilistic \cite{endrenyi2001}.  The models can be continuous such as Poisson or Weibull probabilistic distribution function or composed of deteriorating stages such as in Markov models, or a combination of the two \cite{jiang2004risk}. Failure rate can be assumed either constant, time dependent, or a function of external factors such as weather \cite{mccalley2005computing}. Furthermore, the effect of maintenance actions, which are done to reduce the failure rate of an asset, can be modeled as either deterministic or probabilistic in both the effect of the maintenance and its duration.\\
%\end{appendix}
\end{document}